\documentclass[12pt]{extarticle}
\usepackage{amsmath,amssymb,bm}
\usepackage[T1]{fontenc}
\usepackage{lmodern}
\usepackage{hyperref}
\usepackage{authblk} % <--- Add this

\newcommand{\er}{\hat{\bm e}_r}
\newcommand{\ephi}{\hat{\bm e}_\phi}
\newcommand{\etheta}{\hat{\bm e}_\theta}
\newcommand{\ud}{\,\mathrm d}
\begin{document}
\title{The Vector and Canonical Components of the Momentum Operator in 3D Euclidean Space Spanned by General Curvilinear Coordinates}

\author{M. S. Shikakhwa\thanks{shikakhwa@gmail.com}}
\affil{Department of Basic Sciences, TED University\\ Ziya Gökalp Caddesi No:48, 06420, Ankara, Türkiye}
\affil{and Department of Physics, Middle East Technical University, 06531 Ankara, Türkiye}

%\begin{document}
\maketitle

\begin{abstract}
We construct the Hermitian vector and canonical components of the momentum operator in 3D Euclidean space spanned by general curvilinear coordinates (GCC's) using a simple, natural and unified approach based on identifying the momentum operator in any coordinate system as mass times the velocity operator. When this latter is calculated by applying the Heisenberg equation of motion, it returns  ($-i\hbar$ times) the gradient operator plus an additional zero-valued sum, which when distributed among the components of the gradient, it makes each the Hermitian vector component of the momentum operator in GCC's. The canonical components follow immediately upon symmetrizing each of these vector components in the corresponding base vector. For accessability by wider audiences, we first develop the formalism for the simple polar coordinates and then we develop the case for GCC's. 
\end{abstract}

\maketitle
\section{Introduction}
Discussion of aspects of quantum mechanics in  curvilinear coordinates represents a topic that is
almost absent from most undergraduate and even graduate textbooks \cite{Griffiths,Gaziorowicz,sakurai}. This is only natural, as non-relativistic quantum mechanics in curved coordinate systems, which requires more complex mathematical formulation, has historically
found limited practical applications to justify its discussion in textbooks.  However, recent technological advances, in particular, nanotechnology, has enabled the construction of nano-scale structures such as
quantum balls and quantum wires that inherently define curved geometries. These systems require
a proper quantum mechanical treatment in curved coordinate systems to accurately describe their
physical behavior. Essentially, a helical wire or surface of a spheres are curved structures, and one needs to formulate the Hamiltonians, and in particular the momenta on these geometries using curvilinear coordinates    \cite{shikakhwa_2019physicaE,shikakhwa2018_commTP,shikakhwa2022_EPJPlus,shikakhwa2024_curve, Liu}. Additionally, the quantum mechanical problem of a particle with position-dependent mass can be elegantly reformulated \cite{quesne2004} as that of a constant-mass particle moving in an
appropriately curved space. While these systems live in non-Euclidean curved spaces, an essential prelude to comprehend them is an understanding of the formulation of quantum systems in Euclidean space spanned by general curvilinear coordinates (GCC's). \\ 
Central to formulating quantum dynamics in GCC's is the proper definition of
the momentum operator and its components in these coordinates.  While textbooks routinely discuss the
momentum operator in Cartesian coordinates as $\vec{p} = -i\hbar\nabla$, and identify the components of the Cartesian gradient as the Cartesian components of the momentum, the components of this 
operator in general curvilinear coordinates are rarely addressed comprehensively. For example, in the classic text by Merzbacher  \cite{merzbacher}, there is only a 'shy" mention of the radial canonical momentum showing that it is not just $-i\hbar\frac{\partial}{\partial r}$ as this is not Hermitian. Park's textbook \cite{park} is  among the rare ones that  derives the expression for the canonical momentum in orthogonal GCC's following the same approach of De Witt \cite{DeWitt1,DeWitt2} who was probably the first (see also \cite{Cade}) to construct the general form of the Hermitian canonical momentum in GCC's.  The approach is based on adding a suitable function of coordinates to the derivative so that the resulting operator both satisfies the canonical commutation relations and be Hermitian at the same time  resulting in the expression  
\begin{equation}\label{canonical momentum}
p_n=  -i \hbar\left(\partial_n+\frac{1}{2\sqrt{g}}\frac{\partial\sqrt{g}}{\partial {q^n}}\right)=-i \hbar\left(\partial_n+\frac{1}{2|J|}\frac{\partial|J|}{\partial {q^n}}\right)
\end{equation}
 with $q^n \quad n=1..3$ being the generalized coordinates, $|J|$ the absolute value of the Jacobian of transformation,  and  $|J|=\sqrt{g}$ valid in Euclidean space represents the volume element in general curvilinear coordinates, with $g$ being the determinant of the metric tensor ( see the discussion after Eq.(\ref{metric})) below. This factor ensures that the momentum operator maintains its physical meaning and Hermiticity when moving away from a flat Cartesian grid. The same expression was reached at by other authors following a different approach \cite{gruber1971}. Leaf \cite{Leaf,Leaf FP}  attempted to provide the connection of these canonical momenta to the full momentum vector operator $\vec{p}=-i\hbar\vec{\nabla}$. He \textit{defined} the momentum component operators $p_n$ as   ( $\vec{u}^{\,n}$'s being the base vectors): 
\begin{equation}\label{Leaf definition}
p_n=\frac{1}{2}\left(\vec{p}_{n} \cdot\vec{u}^{\,n}+\vec{u}^{\,n}\cdot\vec{ p}_{n}\right)
\end{equation}
 and expressed the full momentum operator in terms of these components as:
\begin{equation}\label{Leaf2}
 \vec{p}=\frac{1}{2}\left(p_{n}\vec{u}^{\,n}+\vec{u}^{\,n} p_{n}\right)
\end{equation}
with $\vec{p}_n$ and $p_n$ in the first terms of the above equations acting on everything to its right. He then moves to show that the $p_n$'s appearing in his expression above are just the canonical Hermitian momenta given in Eq.(\ref{canonical momentum})derived first by De Witt.\\ 
  
The purpose of the present work is two fold: On the one hand, we introduce a new natural and intuitive approach for constructing the vector and then the scalar canonical components of the momentum operator in GCC's. The approach is based on noting that upon identifying the momentum operator as (mass times ) the velocity operator, then, the dynamics, i.e. the Heisenberg equation of motion, seems to suggest the term that needs to be added to the vector components of the momentum operator to render each Hermitian. Each of the resulting Hermitian vector component is then seen to have the symmetric form of Eq.(\ref{Leaf2}). We also give a physical rationale for extracting the canonical momentum by symmetrically contracting with the base vectors  as in Eq.(\ref{Leaf definition}), which was not clarified by Leaf. On the other hand, the work represents a single resource, which is not available in the literature, where both the vector and scalar components of momentum in GCC's are constructed in a single approach, so learners of quantum mechanics can find the whole thing consistently presented in a single place. For pedagogical purposes and to make the presentation accessible to learners of diverse mathematical background, we start by applying the formalism to Polar coordinates, the simplest of curvilinear coordinates, in section 2,  then move to the general case of GCC's in section 3.

 \section{Hermitian Vector and Canonical Momentum Operator Components in Polar Coordinates}  
Let $(q^1,q^2)=(r,\phi)$ be plane polar coordinates (PC's) on $\mathbb R^2$, with
\begin{equation}
x=r\cos\phi,\qquad y=r\sin\phi,
\qquad r\in[0,\infty),\ \phi\in[0,2\pi).
\end{equation}
Write $\vec r=x\,\hat{\bm e}_x+y\,\hat{\bm e}_y=r\,\er$, where $\er,\ephi$ are the standard polar unit vectors and
\begin{equation}
\partial_\phi \er=\ephi,\qquad \partial_\phi \ephi=-\er.
\end{equation}
Since the discussion of PC's is a prelude to the more general GCC's, we will also introduce and work with the covariant base vectors \cite{auckland}  defined by $\vec u_i=\partial_i\vec r,\quad i=r,\phi$, giving
\begin{equation}\label{metric}
\vec u_r=\er,\qquad \vec u_\phi=r\,\ephi.
\end{equation}
Hence the metric is $g_{ij} =
\begin{pmatrix}
1 & 0 \\
0 & r^2
\end{pmatrix}$, and
\begin{equation}\nonumber
g=\det(g_{ij})=r^2,\qquad \sqrt g=r,
\qquad
(g^{ij})=\begin{pmatrix}
1 & 0 \\
0 & \frac{1}{r^2}
\end{pmatrix}.
\end{equation}
The matrix $(g_{ij})$ introduced in above is the metric tensor. To understand its physical role, consider the infinitesimal distance $ds^2$ in Cartesian coordinates where all coordinates $(x, y, z)$ are lengths, resulting in $ds^2 = dx^2 + dy^2 + dz^2$. However, in curvilinear systems such as polar coordinates $(r, \phi)$, the coordinates are not dimensionally uniform; while $dr$ is a length, $d\phi$ is a dimensionless angle. 
To ensure $ds^2$ maintains the consistent dimensions of $(\text{length})^2$, we must multiply $d\phi$ by a scaling factor with dimensions of length (in this case, $r$), leading to the familiar form $ds^2 = (1)dr^2 + (r^2)d\phi^2$. Generally, we write this as $ds^2 = g_{ij}dq^i dq^j$, where the metric tensor $g_{ij}$ provides the necessary conversion factors to relate coordinate changes to physical distances. 
The determinant of this tensor, $g$, is significant because its square root, $\sqrt{g}$, acts as a local scaling factor for volumes. In any transformation from Cartesian to curvilinear coordinates, the relationship $\sqrt{g} = |J|$ holds, where $J$ is the Jacobian determinant. Geometrically, this identity reveals that $\sqrt{g}$ represents the ratio of the volume of an infinitesimal "box" in curvilinear space to the corresponding volume in Cartesian space. As discussed in standard mathematical physics texts \cite{Boas,Arfken},this factor is essential for defining an invariant physical volume $dV = \sqrt{g} \, dq^1 dq^2 dq^3$ that remains independent of the specific coordinate system chosen.
The inner product is therefore
\begin{equation}
\langle\psi|\chi\rangle=\int_0^\infty\!\ud r\int_0^{2\pi}\!\ud\phi\; \sqrt{g}\,\psi^*(r,\phi)\,\chi(r,\phi)=\int_0^\infty\!\ud r\int_0^{2\pi}\!\ud\phi\; r\,\psi^*(r,\phi)\,\chi(r,\phi).
\end{equation}
The contravariant base vectors $\vec u^{\,i}=g^{ij}\vec u_j$ are
\begin{equation}\label{cv polar base vectors}
\vec u^{\,r}=\er,\qquad
\vec u^{\,\phi}=\frac1r\,\ephi.
\end{equation}

The momentum operator in any coordinate system is $-i\hbar$ times the gradient in these coordinates. With the gradient in GCC's being $\vec u^{\,i}\partial_i$, this gives in PC's the expected form :
\begin{equation}\label{6}
\vec p=-i\hbar\,\vec u^{\,i}\partial_i
=-i\hbar\left(\vec u^{\,r}\partial_r+\vec u^{\,\phi}\partial_\phi\right)
=-i\hbar\left(\er\,\partial_r+\frac1r\,\ephi\,\partial_\phi\right).
\end{equation}

Now, if we define the vector components of this momentum operators na\"ively following what we do in Cartesian coordinates, we have: 
\[
\vec p_r:=-i\hbar\,\vec u^{\,r}\partial_r=-i\hbar\,\er\,\partial_r,
\qquad
\vec {p}_\phi:=-i\hbar\,\vec u^{\,\phi}\partial_\phi=-i\hbar\,\frac1r\,\ephi\,\partial_\phi.
\]
Both of these components, however, fail the check of Hermicity as we show explicitly now. Hermicity check starts from  :  
\begin{equation}
\langle\psi|\vec {p}_i\psi\rangle
=\int_0^\infty\!\ud r\int_0^{2\pi}\!\ud\phi\; r\,\psi^*\Big(-i\hbar\,\vec u^{\,i}\partial_i\psi\Big),
\qquad i=r,\phi .
%\tag{7b}
\end{equation}

Using $\vec {p}_\phi=-i\hbar\,\frac1r\,\ephi\,\partial_\phi$, note that the measure factor $r$ cancels the $1/r$, so
\[
\langle\psi|\vec {p}_\phi\psi\rangle
=\int_0^\infty\!\ud r\int_0^{2\pi}\!\ud\phi\;\psi^*\Big(-i\hbar\,\ephi\,\partial_\phi\psi\Big).
\]
Integrate by parts in $\phi$ :
\begin{align}
\int_0^{2\pi}\!\ud\phi\;\psi^*(-i\hbar\,\ephi\,\partial_\phi\psi)
&=
\Big[-i\hbar\,\psi^*\,\ephi\,\psi\Big]_{0}^{2\pi}
+\int_0^{2\pi}\!\ud\phi\;i\hbar\,(\partial_\phi\psi^*)\,\ephi\,\psi\\\nonumber
&+\int_0^{2\pi}\!\ud\phi\;i\hbar\,\psi^*(\partial_\phi\ephi)\psi.
\end{align}
The surface term vanishes for periodic/single-valued boundary conditions in $\phi$:
$\psi(r,2\pi)=\psi(r,0)$ and $\ephi(2\pi)=\ephi(0)$. Thus,
\begin{equation}\label{non-Hermitian vector phi}
\langle\psi|\vec {p}_\phi\psi\rangle
=\langle \vec {p}_\phi\psi|\psi\rangle
+\left\langle -i\hbar\,\frac{1}{\sqrt g}\,\partial_\phi\big(\sqrt g\,\vec u^{\,\phi}\big)\psi\ \middle|\ \psi\right\rangle,
\qquad \sqrt g=r.
%\tag{7c$_\phi$}
\end{equation}
Since $\sqrt g\,\vec u^{\,\phi}=r\cdot\frac1r\,\ephi=\ephi$,
\[
\frac{1}{\sqrt g}\,\partial_\phi(\sqrt g\,\vec u^{\,\phi})
=\frac1r\,\partial_\phi(\ephi)
=\frac1r(-\er)\neq 0,
\]
so $\vec {p}_\phi$ is not Hermitian. The same applies for the $r$-component

\[
\langle\psi|\vec p_r\psi\rangle
=\int_0^\infty\!\ud r\int_0^{2\pi}\!\ud\phi\; r\,\psi^*\Big(-i\hbar\,\er\,\partial_r\psi\Big).
\]
Integrate by parts in $r$ :
\begin{align}
\int_0^\infty\!\ud r\; r\,\psi^*\,\er\,\partial_r\psi
&=
\Big[r\,\psi^*\,\er\,\psi\Big]_{0}^{\infty}
-\int_0^\infty\!\ud r\;\partial_r\!\big(r\,\psi^*\,\er\big)\,\psi.
\end{align}
The boundary term is dropped under standard physical conditions (square-integrability/decay as $r\to\infty$, and regularity at $r=0$ ensuring $r\,\psi^*\psi\to0$). One then obtains
\begin{equation}\label{non-Hermitian vector r}
\langle\psi|\vec p_r\psi\rangle
=\langle \vec p_r\psi|\psi\rangle
+\left\langle -i\hbar\,\frac{1}{\sqrt g}\,\partial_r\big(\sqrt g\,\vec u^{\,r}\big)\psi\ \middle|\ \psi\right\rangle,
\qquad \sqrt g=r.
%\tag{7c$_r$}
\end{equation}
Since $\sqrt g\,\vec u^{\,r}=r\,\er$,
\[
\frac{1}{\sqrt g}\,\partial_r(\sqrt g\,\vec u^{\,r})
=\frac1r\,\partial_r(r\,\er)=\frac1r\,\er\neq 0,
\]
so $\vec p_r$ is not Hermitian.

It is similarly easy to check that the na\"ive scalar coordinate-derivative operators are problematic: in particular $-i\hbar\,\partial_r$ is not Hermitian under the weight $r$. Therefore, this cannot be taken as the canonical momentum conjugate to $r$. In polar coordinates, $-i\hbar\,\partial_\phi$ is Hermitian under periodicity, but this does \emph{not} cure the non-Hermiticity of the \emph{vector} $\phi$-piece, which fails because the basis depends on $\phi$.

For the \emph{full} momentum operator $\vec p=\vec p_r+\vec {p}_\phi$, the same integration-by-parts logic yields
\begin{equation}
\langle\psi|\vec p\,\psi\rangle
=\langle \vec p\,\psi|\psi\rangle
+\left\langle -i\hbar\,\frac{1}{\sqrt g}\,\partial_i(\sqrt g\,\vec u^{\,i})\psi\ \middle|\ \psi\right\rangle,
\qquad i=r,\phi.
%\tag{8}
\end{equation}
Expanding the sum over $i$ in the second term we get,
\[
\frac{1}{\sqrt g}\,\partial_i(\sqrt g\,\vec u^{\,i})
=
\frac{1}{\sqrt g}\Big[\partial_r(\sqrt g\,\vec u^{\,r})+\partial_\phi(\sqrt g\,\vec u^{\,\phi})\Big]
=\frac1r\Big[\partial_r(r\er)+\partial_\phi(\ephi)\Big].
\]
But $\partial_r(r\er)=\er$ and $\partial_\phi(\ephi)=-\er$, so
\begin{equation}\label{vanishing sum polar}
\partial_r(r\er)+\partial_\phi(\ephi)= 0
\quad\Longrightarrow\quad
\frac{1}{\sqrt g}\,\partial_i(\sqrt g\,\vec u^{\,i})= 0.
%\tag{9}
\end{equation}
Therefore the ``Hermicity breaking term'' cancels in the sum, and the full momentum operator given by Eq.(\ref{6}) is Hermitian.\footnote{Strictly speaking, we should  denote an operator $A$ which satisfies the condition $\langle A \psi \mid \psi\rangle=\langle \psi \mid A\psi\rangle$ symmetric, as having an operator Hermitian or self-adjoint takes more than just satisfying this relation, see \cite{Dennery}, for example. The earlier terminology is widely used and became standard, however, and we use it in this work.} 

Now, we move to introduce an approach that will enable us to find the Hermitian vector components of the full momentum in a natural way. Let us recall that with a standard Hamiltonian with velocity-independent scalar potential only, the momentum operator is mass times the velocity operator. This means that we can identify it as $\vec{p}=m\,\dot{\vec r}$
by calculating the time derivative of the position operator using the Heisenberg equation of motion. 
The Hamiltonian with the Laplacian expressed as in the Laplace-Beltrami form is: 
\[
H=-\frac{\hbar^2}{2m}\,\Delta+V(\vec r),
\qquad
\Delta=\frac{1}{\sqrt g}\partial_i\big(\sqrt g\,g^{ij}\partial_j\big),
\]
 and the Heisenberg equation is:
\[
m\frac{d\vec r}{dt}=\frac{m}{i\hbar}[\vec r,H].
\]
For polar coordinates, the Laplacian has the explicit form:
\begin{equation}\label{polar laplacian}
\nabla^{2}
=\frac{1}{r}\,\partial_{r}\!\left(r\,\partial_{r}\right)
+\frac{1}{r^{2}}\,\partial_{\phi}^{2}
\end{equation}
A direct evaluation of the commutator (see Appendix A.1) yields, in polar coordinates,
\begin{equation}\label{mv polar}
m\frac{d\vec r}{dt}
=-i\hbar\left[
\vec u^{\,r}\partial_r+\vec u^{\,\phi}\partial_\phi
+\frac{1}{2\sqrt g}\Big(\partial_r(\sqrt g\,\vec u^{\,r})+\partial_\phi(\sqrt g\,\vec u^{\,\phi})\Big)
\right].
%\tag{10}
\end{equation}
The first two terms together give the momentum operator,  Eq.(\ref{6}), and the sum of the last two terms  is just  \emph{half} the zero sum given by Eq.(\ref{vanishing sum polar}),  so $m\dot{\vec r}=\vec p$ as expected. Here we note the interesting point that Eq.(\ref{mv polar}) naturally splits into two coordinate contributions  :
\begin{equation}\label{Hermitian p polar}
\vec{p}=-i\hbar\left(\vec u^{\,r}\partial_r+\frac{1}{2\sqrt g}\partial_r(\sqrt g\,\vec u^{\,r})\right)
\;+\;
-i\hbar\left(\vec u^{\,\phi}\partial_\phi+\frac{1}{2\sqrt g}\partial_\phi(\sqrt g\,\vec u^{\,\phi})\right).
\end{equation}
where the non-derivative part in each term is not vanishing by itself as we have shown before. Comparing with Eqs.(\ref{non-Hermitian vector phi}) and (\ref{non-Hermitian vector r}), the dynamics suggests adding \emph{one half} of the respective Hermicity-breaking term to each separated vector component! This, interestingly,  renders each piece Hermitian, as it cancels half the breaking term, while the full momentum remains unchanged because the sum of the half-terms is one half of an identically vanishing quantity.

Accordingly, we define the Hermitian vector components (no sum on $i$):
\begin{equation}\label{polar vector p's lumped}
\vec p_i^{\,H}:=-i\hbar\left(\vec u^{\,i}\partial_i+\frac{1}{2\sqrt g}\partial_i(\sqrt g\,\vec u^{\,i})\right),
\qquad i=r,\phi,\qquad \sqrt g=r.
\end{equation}
or, expressed in terms of unit vectors: 
\begin{equation}\label{polar vector p's}
\vec p_{r}^{\,H}= -\mathrm i\hbar\!\left(\er\,\partial_{r}+ \frac{1}{2r}\,\er\right), 
\qquad
\vec p_{\phi}^{\,H}= -\mathrm i\hbar\!\left(\frac{\ephi}{r}\,\partial_{\phi}- \frac{1}{2r}\,\er\right).
\end{equation}
 Summing Eq.(\ref{polar vector p's lumped}) gives
\begin{eqnarray}
\vec p^{\,H}&=&\vec p_r^{\,H}+\vec p_\phi^{\,H}
=-i\hbar\left(\vec u^{\,r}\partial_r+\vec u^{\,\phi}\partial_\phi+\frac{1}{2\sqrt g}\partial_i(\sqrt g\,\vec u^{\,i})\right)\\\nonumber
&=&-i\hbar\left(\er\,\partial_r+\frac1r\,\ephi\,\partial_\phi\right)=\vec p.
\end{eqnarray}
since $\partial_i(\sqrt g\,\vec u^{\,i})=\ 0$. 
The appearance of the unit vector $\er$ in the expression for $\vec p_{\phi}^{\,H}$ in Eq.(\ref{polar vector p's}) is a bit unusual, given the expressions we are used to in the Cartesian vector components of $\vec{p}$, and might lead one to think that $\vec p_{\phi}^{\,H}$ has a component along $\er$. Of course, this is not the case, and this situation raises  the question of how to determine the components of $\vec{p}$, to which we now turn. We also note that  we encounter similar situation  for the Hermitian vector components in GCC's as we will show in the next section. To construct the components, we start from Eq.(\ref{polar vector p's lumped}) and sum it over $i=r,\phi$,
\begin{equation}\nonumber
\frac{1}{2\sqrt g}\partial_i(\sqrt g\,\vec u^{\,i})
=\frac12\,\partial_i\vec u^{\,i}+\frac12(\partial_i\ln\sqrt g)\,\vec u^{\,i}.
\end{equation}
where we have noted that $\frac{1}{2\sqrt g}\partial_i(\sqrt g)=\frac12(\partial_i\ln\sqrt g)$. Define
\[
D_i:=\partial_i+\frac12\,\partial_i\ln\sqrt g.
\]
Then Eq.(\ref{polar vector p's lumped}) can be written in the manifestly symmetric form
\begin{equation}\label{symmetric polar}
\vec p_i^{\,H}
=-\frac{i\hbar}{2}\Big(\vec u^{\,i}D_i + D_i\vec u^{\,i}\Big),
\qquad i=r,\phi.
\end{equation}
For polar coordinates, $\sqrt g=r$ so $\partial_r\ln\sqrt g=1/r$ and $\partial_\phi\ln\sqrt g=0$, hence
\begin{equation}\label{canonical polar p}
D_r=\partial_r+\frac{1}{2r},\qquad D_\phi=\partial_\phi.
\end{equation}
Using $\vec u^{\,r}=\er$ and $\vec u^{\,\phi}=\frac1r\,\ephi$, we obtain the symmetric unit-vector forms ( note that the derivatives $\partial_r$ and $\partial_\phi$ act on everything to their right including the unit vectors):
\begin{equation}\label{symmetric p_r}
\vec p_r^{\,H}
=-\frac{i\hbar}{2}\left[\er\Big(\partial_r+\frac{1}{2r}\Big)+\Big(\partial_r+\frac{1}{2r}\Big)\er\right],
\end{equation}
\begin{equation}\label{symmetric p_phi}
\vec p_\phi^{\,H}
=-\frac{i\hbar}{2}\left[\frac1r\,\ephi\,\partial_\phi+\partial_\phi\!\Big(\frac1r\,\ephi\Big)\right].
\end{equation}

The symmetric structure in Eqs.(\ref{symmetric p_r}) and (\ref{symmetric p_phi})  naturally suggests identifying the canonical scalar momenta
\begin{equation}\label{canonical polar p's}
p_r:=-i\hbar\left(\partial_r+\frac{1}{2r}\right),
\qquad
p_\phi:=-i\hbar\,\partial_\phi,
\end{equation}
since these are precisely the scalar operators appearing inside the symmetrization and are also Hermitian as one can easily check. Note also that the part of $\vec p_\phi^{\,H}$ that appears multiplying $\er$ in Eq.(\ref{polar vector p's}) does not contribute to the canonical component $p_\phi$ given above. Of course, we still need to show how these components can be projected from the vector expressions by dotting with the respective base vectors. Before doing this, we need to note that since the latter are generally functions of the coordinates, we need to symmetrize as we project. 
%\begin{equation}
%\boxed{
%\vec p_r^{\,H}=\frac12\big(\vec u^{\,r}p_r+p_r\vec u^{\,r}\big),
%\qquad
%\vec p_\phi^{\,H}=\frac12\big(\vec u^{\,\phi}p_\phi+p_\phi\vec u^{\,\phi}\big),
%\qquad
%\vec p=\vec p_r^{\,H}+\vec p_\phi^{\,H}.
%}
%\tag{20}
%\end{equation}
Using $\vec u_r=\er$ and $\vec u_\phi=r\,\ephi$, define the $r$ and $\phi$ components as 
\begin{equation}\label{projected canonical polar p's}
p_r=\frac12\big(\vec u_r\cdot\vec p_r^{\,H}+\vec p_r^{\,H}\cdot\vec u_r\big),
\qquad
p_\phi=\frac12\big(\vec u_\phi\cdot\vec p_\phi^{\,H}+\vec p_\phi^{\,H}\cdot\vec u_\phi\big),
\end{equation}
which can be easily seen to reproduce exactly the components  in Eq.(\ref{canonical polar p's}).
Finally, the momenta defined in Eq.(\ref{canonical polar p's}) satisfy the canonical commutation relations:
\[
[r,p_r]=i\hbar,\qquad [\phi,p_\phi]=i\hbar,\qquad [r,p_\phi]=[\phi,p_r]=0,
\]
confirming their role as the canonical momenta conjugate to $(r,\phi)$. 

We have now a well-defined straightforward algorithm to construct the Hermitian vector and canonical momentum operators that - as we will show in the next section- applies to any coordinate system spanning the 3D Euclidean space: First, find ( mass times) the velocity operator using Heisenberg equations, then bring this to the form given in Eq.(\ref{mv polar}), thus define the vector momentum components as per Eq.(\ref{polar vector p's lumped}), and finally project the Hermitian canonical components as per Eq.(\ref{projected canonical polar p's}). We leave it as an exercise to the reader to apply this algorithm to spherical polar coordinates. The answers are given in Appendix A.2

\section{Momentum operator in General Curvilinear Coordinates}
In this section, we generalize our approach to GCC's and derive expressions for the Hermitian components of momentum that are valid in any curvilinear coordinate system. In Euclidean space \(\mathbb{E}^3\), a general curvilinear coordinate system \(\{q^a\}\) is given by the invertible transformation
\begin{equation}
x^m = x^m(q^1,q^2,q^3), 
\quad
q^a = q^a(x^1,x^2,x^3),
\quad m,a = 1,2,3,
\end{equation}
$x^m$'s are the Cartesian coordinates, with the usual Jacobian relations
\begin{equation}
\frac{\partial x^m}{\partial q^a}\,\frac{\partial q^a}{\partial x^n}
=\delta^m_n,
\quad
\frac{\partial q^a}{\partial x^m}\,\frac{\partial x^m}{\partial q^b}
=\delta^a_b.
\end{equation}
The position vector is given in Cartesian coordinates by the well-known expression: 
\[
\vec r(x^m)=x^m\,\hat e_m,
\]
with $\hat e_m=\frac{\partial\vec r}{\partial x^m}$ being the Cartesian unit vectors. The base vectors and metric tensor of GCC's are given by
\[
\vec u_a=\frac{\partial\vec r}{\partial q^a},
\qquad
g_{ab}=\vec u_a\cdot\vec u_b.
\]
So, the position vector can also be expressed as:
\[
\vec r(q^a)=x^m(q^a)\,\frac{\partial\vec r}{\partial q^a}\frac{\partial q^a}{\partial x^m}=(x^m(q^a)\frac{\partial q^a}{\partial x^m})\vec u_a
\]
 We use the following conventions throughout the rest of the article: $\partial_{i} \equiv \frac{\partial}{\partial q^{i}}$, when we want to refer to Cartesian coordinates or derivatives with respect to these, we will write that explicitly. Also, Einstein summation convention over repeated indices will be adopted when we are using  letters from the middle of the alphabets; $i, j, k$ ..etc. However, when we use letters from the beginning of the alphabets; $ a,b,c$...etc. no summation is implied for repeated indices unless we explicitly introduce summation, i.e. $\sum_a$; say.\\
 The momentum operator,  expressed as usual in terms of the gradient, takes in GCC's the form ($\vec{u}^i=g^{ij}\vec{u}_j$ with $\vec{u}_j$ the base vectors defined above):
  \begin{equation}\label{}
\vec{p}=-i \hbar \vec{\nabla}=-i \hbar \vec{u}^i \partial_{i}
  \end{equation} 
  This operator is, as we have noted before,  necessarily Hermitian regardless of the coordinate system we use, as it is just the gradient operator. However,  the momentum component vectors $\vec{p}_{a}=-i \hbar \vec{u}^a \partial_a$ (no sum) are not Hermitian :
\begin{equation}\label{not hermitian}
\begin{aligned}
& \left\langle\psi \mid \vec{p}_a \psi\right\rangle=\int d^{3} q \sqrt{g} \psi^{*}\left(-i \hbar \vec{u}^a \partial_a \psi\right) \\
& =-i \hbar \int d^{3} q \partial_{a}\left(\sqrt{g} \psi^{*} \vec{u}^a \psi\right) 
 +\int d^{3} q \sqrt{g}\left(-i \hbar \vec{u}^a \partial_a \psi\right)^{*} \psi+ 
 \int d^{3}q  \sqrt{g}\left(\frac{-i \hbar}{\sqrt{g}}\partial_a\left(\sqrt{g} \vec{u}^a\right) \psi\right)^{*} \psi \\
& =\left\langle\vec{p}_a \psi \mid \psi\right\rangle+\langle\frac{-i \hbar}{\sqrt{g}} \partial_a \left(\sqrt{g} \vec{u}^a\right) \psi|\psi\rangle
\end{aligned}
\end{equation}

$a=1, \cdots 3$, $\sqrt{g}$ is the determinant of the metric tensor which is at the same time (when multiplied by $d^3 q$ ) is the volume element in GCC's, and we have dropped the surface term by imposing the standard boundary conditions on the wave function at the surface. The term $\frac{-i \hbar}{\sqrt{g}} \partial_a \left(\sqrt{g} \vec{u}^a\right) $ can be expressed in terms of the Christoffel symbols of the second kind (see below) and is,  generally, non-vanishing. One can further check that taking $p_a=-i\hbar\partial_a$ as the component of the momentum along $\vec{u}_a$  does not work either, since this turns out to be not Hermitian for the same reason.\\
Eq. (\ref{not hermitian}) suggests that, for the full momentum operator $\vec{p}=-i \hbar \vec{u}^{i} \partial_{i}$, we will get (note the repeated index is summed over here):

\begin{equation}\label{not hermitian 2}
\langle\psi \mid \vec{p} \psi\rangle = \langle\vec{p} \psi \mid \psi\rangle + \langle -\frac{i \hbar}{\sqrt{g}} \partial_{i}\left(\sqrt{g} \vec{u}^{i}\right) \psi \mid \psi \rangle
\end{equation}
The second term in Eq. (\ref{not hermitian 2}) which is a generalization of Eq.(\ref{vanishing sum polar}) in PC's differs from the corresponding term in Eq. (\ref{not hermitian}) in that we have summation over $i$ in Eq. (\ref{not hermitian 2}) which make it vanish identically:
\begin{equation}\label{zero identity}
\begin{aligned}
\frac{1}{\sqrt{g}} \partial_{i}\left(\sqrt{g} \vec{u}^{i}\right)&=\frac{1}{\sqrt{g}}\left(\partial_{i} \sqrt{g}\right) \vec{u}^{i}+\partial_{i} \vec{u}^{i}\\
&=\Gamma_{i j}^{j} \vec{u}^{i}-\Gamma_{i j}^{i} \vec{u}^{j}=0 .
\end{aligned}
\end{equation}

where we have used the well-known identities \cite{auckland}: 
\begin{equation}\label{CS identities}
\frac{1}{\sqrt{g}} \partial_{i}(\sqrt{g})=\Gamma_{i j}^{j} ,\qquad \partial_{i} \vec{u}^{i}=-\Gamma_{ij}^{i} \vec{u}^{j}
\end{equation} . 
Here $\Gamma_{i j}^{k}$ are the Christoffel symbols of the second kind that are symmetric in the lower indices $\Gamma_{i j}^{k}=\Gamma_{ ji}^{k}$, and vanish when all indices are the same. The readers who are not familiar with these symbols just need to view these functions of the generalized coordinates as the expansion coefficients of the derivatives of a  base vector along the base vectors of the space. We will need only the above two identities of these symbols in almost all of our manipulations in this section. 
 The vanishing of the second term in Eq. (\ref{not hermitian 2}) is, of course, a necessity since the momentum operator $\vec{p}$ should be Hermitian in any coordinate system as we have noted before. \\
  We now proceed as in the PC's case to construct the Hermitian vectors components of the momentum by calculating $m \frac{d \vec{r}}{d t}=m\vec{v}$ using the Heisenberg equation of motion.   Applying the Heisenberg equation of motion with 
 $$
 H=\frac{-\hbar^2}{2 m} \nabla^{2}+V(\vec{r})=\frac{-\hbar^2}{2 m}\left(\frac{1}{\sqrt{g}} \partial_{i} \sqrt{g} g^{i j} \partial_{j}\right)+V(\vec{r})
 $$  we have:

\begin{align}\label{eq:motion}
m \frac{d \vec{r}}{d t} &= \frac{1 }{i \hbar}[\vec{r}, H] = \frac{i \hbar}{2 }\left([\vec{r}, \frac{1}{\sqrt{g}} \partial_{i}\left(\sqrt{g} g^{i j}\right) \partial_{j}] + \left[\vec{r}, g^{i j} \partial_{i} \partial_{j}\right]\right) \\\nonumber
&= -i \hbar\left(\vec{u}^{i} \partial_{i} + \frac{1}{2 \sqrt{g}} \partial_{i} \left(\sqrt{g} \vec{u}^{i}\right)\right)=\vec{p}.
\end{align}

Again, as in PC's, $\frac{m d \vec{r}}{d t}=m \vec{v}$  returns  the gradient plus  \emph{half} the zero-valued 
  $\frac{1}{2\sqrt{g}} \partial_{i}\left(\sqrt{g} \vec{u}^{i}\right)$ term. When we add this zero-valued term to each $\vec{p}_a=-i \hbar \vec{u}^{a} \partial_a$, we can readily see, for the same reason as in PC's ( see the paragraph following Eq.(\ref{Hermitian p polar})) that the resulting operators are  Hermitian:

\begin{equation}\label{vector p hermitian}
\vec{p}_a^{\,H}=-i \hbar\left(\vec{u}^a \partial_a+\frac{1}{2 \sqrt{g}} \partial_a\left(\sqrt{g} \vec{u}^a\right)\right), a=1 \ldots 3
\end{equation}
The momentum operator that we construct by adding all the $\vec{p}_a^{\,H} $'s is still $\vec{p}=-i \hbar \vec{u}^{i} \partial_{i}$ since $\frac{1}{2 \sqrt{g}} \partial_{i}\left(\sqrt{g} \tilde{u}^{i}\right)=0$ (recall, we are summing over $i$  here).So, we write \cite{shikakhwa_2019physicaE,shikakhwa2018_commTP}:

%\begin{equation}\label{hermitian gradient}
%\vec{p}=-i \hbar\left(\vec{u}^{i} \partial_{i}+\frac{1}{2 \sqrt{g}} \partial_{i}\left(\sqrt{g} \vec{u}^{i}\right)\right)=-i \hbar\vec{u}^{i} \partial_{i}
%\end{equation}
\begin{equation}\label{}
 \vec{p}=m \frac{d \vec{r}}{d t}=\sum_{a}^{3} \vec{p}_a^{\,H}
\end{equation} 
Before moving on to construct the canonical components of the momentum, we examine closely these  component  vector operators $\vec{p}_a^{\,H}$ that we have constructed in Eq. (\ref{vector p hermitian}) which, while Hermitian, are - as we have noted in PC's- a bit strange in comparison to those in Cartesian coordinates. Let us set  $a=1$, for example, then

\begin{equation}\label{weird p}
\begin{aligned}
& \quad \vec{p}_1^{\,H}=-i \hbar\left(\vec{u}^{1} \partial_{1}+\frac{1}{2 \sqrt{g}} \partial_{1}\left(\sqrt{g} \vec{u}^{1}\right)\right) \\
& \quad=-i \hbar\left(\vec{u}^{1} \partial_{1}+\frac{1}{2 \sqrt{g}} \partial_{1}(\sqrt{g}) \vec{u}^{1}+\frac{1}{2} \partial_{1} \vec{u}^{1}\right) \\
& \quad=-i \hbar\left(\vec{u}^{1} \partial_{1}+\frac{1}{2} \Gamma_{1 k}^{k} \vec{u}^{1}-\frac{1}{2} \Gamma_{1 k}^{1} \vec{u}^{k}\right) \\
& =-i \hbar\left(\vec{u}^{1}\left(\partial_{1}+\frac{1}{2}\left(\Gamma_{12}^{2}+\Gamma_{13}^{3}\right)\right)-\frac{1}{2} \Gamma_{12}^{1} \vec{u}^{2}\right. \\
& \left.\quad-\frac{1}{2} \Gamma_{13}^{1} \vec{u}^{3}\right)
\end{aligned}
\end{equation}

where we have used the Christoffel symbols identities given in  Eq.(\ref{CS identities}) in the derivation. This $\vec{p}_1^{\,H}$ seems quite unusual, especially, when we - driven by what we know from Cartesian components of the momentum - expect to have "some derivative" multiplying the base vector $\vec{u}^{1}$ only; but not the appearance of $\vec{u}^{2}$ and $\vec{u}^{3}$ base vectors. Of course, we have, when working in GCC's, to redefine what we understand when we say the component of a differential vector  operator. Let us see how we can define components  of $\vec{p}$. In fact, as in PC's, the strange form of $\vec{p}_1^{\,H}$ is showing us the way: The second line of Eq. (\ref{weird p}) suggests a symmetrization of $\vec{p}_{1}^{\,H}$ as:
\begin{equation}\label{p_1 hermitian}
\begin{aligned}
\vec{p}_1^{\,H}=&-i \hbar\left(\vec{u}^{1} \partial_{1}+\frac{1}{2 \sqrt{g}} \partial_{1}\left(\sqrt{g} \vec{u}^{1}\right)\right)=-i \hbar\left(\vec{u}^{1} \partial_{1}+\frac{1}{2 \sqrt{g}} \partial_{1}(\sqrt{g}) \vec{u}^{1}+\frac{1}{2} \partial_{1} \vec{u}^{1}\right) \\
=&-\frac{i \hbar}{2}\left(\vec{u}^{1}\left(\partial_{1}+\frac{\Gamma_{1 k}^{k}}{2}\right)+\left(\partial_{1}+\frac{\Gamma_{1 k}^{k}}{2}\right) \vec{u}^{1}\right)
\end{aligned}
\end{equation}
 where we have noted that $\frac{\Gamma_{1 k}^{k}}{2}=\frac{1}{2 \sqrt{g}} \partial_{1}(\sqrt{g})$, and in the second term  the differential operator is taken to act on \textit{everything to its right}, not only on $\vec{u}^{1}$, i.e.
  $$\partial_{1} \vec{u}^{1}=\left(\partial_{1} \vec{u}^{1}\right)+\vec{u}^{1} \partial_{1}=-\Gamma_{1 k}^{1}\vec{u}^{k}+\vec{u}^{1} \partial_{1}
  $$
It seems that, as we did in PC's, identifying $-i\hbar\left(\partial_{1}+\frac{\Gamma_{1 k}^{k}}{2}\right)=-i\hbar\left(\partial_{1}+\frac{1}{2 \sqrt{g}} \partial_{1}(\sqrt{g})\right)$ as $p_{1}^{H}$; the scalar component of $\vec{p}_1^{\,H}$ and so the canonical component conjugate to $q^1$ makes perfect sense now, as it is what multiplies the respective base vectors in a symmetrical manner as per  the general approach used in GCC's \cite{DeWitt1, DeWitt2,Leaf}.
It is straightforward (see below) to see that one can project the components $p_{1}^{H}$ from $\vec{p}_1^{\,H}$ and $\vec{p}$ by symmetrically dotting with the base vectors $\vec{u}_{1}$ :
$$p_{1}^{H}=:\left(\partial_{1}+\frac{\Gamma_{1k}^{k}}{2}\right) =\frac{1}{2}\left(\vec{u}_{1} \cdot \vec{p}^H_1+\vec{p}^H_1 \cdot \vec{u}_{1}\right)=\frac{1}{2}\left(\vec{u}_{1} \cdot \vec{p}+\vec{p} \cdot \vec{u}_{1}\right)
$$
where $\vec{p}=-i \hbar \vec{u}^{i} \partial_{i}$, and again $\vec{p}$ and  $\vec{p}^H_1$  in the second terms  act on everything to the right. The above  defines how we construct the components of $\vec{p}$. Generalizing to all $a=1..3$, we have now the general results:
 \begin{equation}\label{symmetric vector p_a}
 \begin{aligned}
 \vec{p}_a^{\,H}&=-i \hbar\left(\vec{u}^a \partial_a+\frac{1}{2 \sqrt{g}} \partial_a\left(\sqrt{g} \vec{u}^a\right)\right)\\
 &=\frac{1}{2}\left(p_a^{H} \vec{u}^a+\vec{u}^a p_a^{H}\right)
 \end{aligned}
\end{equation}
with,
\begin{subequations}\label{P_a Hermitian}
\begin{align}
 p_a^{H}= &-i\hbar\left(\partial_{a}+\frac{1}{2 \sqrt{g}} \partial_{a}(\sqrt{g})\right)= -i \hbar\left(\partial_a+\frac{\Gamma_{a k}^{k}}{2}\right) \label{P_a Hermitiana}\\
    =&\frac{1}{2}\left(\vec{u}_a \cdot \vec{p}_a^{\,H}+\vec{p}_a^{\,H}\cdot \vec{u}_a\right)\label{P_a Hermitianb} \\
   =&\frac{1}{2}\left(\vec{u}_a \cdot \vec{p}+\vec{p}\cdot \vec{u}_a\right) \label{P_a Hermitianc}
\end{align}
\end{subequations}

and so,

\begin{equation}\label{symmetric gradient}
\vec{p}=-i\hbar\vec{\nabla}=-i \hbar\left(\vec{u}^{i} \partial_{i}+\frac{1}{2 \sqrt{g}} \partial_{i}\left(\sqrt{g} \vec{u}^{i}\right)\right)=\frac{1}{2}\left(p_{i}^H \vec{u}^{i}+\vec{u}^{i} p_{i}^H\right).
\end{equation}
One can verify Eqs.(\ref{P_a Hermitianb}) and (\ref{P_a Hermitianc}) upon using the result derived in the Appendix:
\begin{equation}\label{appendix result}
p_a^{H}=\frac{1}{2}\left(\vec{u}_a \cdot \vec{p}_b+\vec{p}_b\cdot \vec{u}_a\right)
\end{equation}
The first of Eqs.(\ref{P_a Hermitian}) was first reported by De Witt, and its projection as in the second and third line of this equation by Leaf ( Eqs.(\ref{canonical momentum}) and (\ref{Leaf definition})in the Introduction). They are derived here, however,  within a new single natural approach which shows that the Hermitian vector components follow from the construction of  ( mass times) the velocity operator as in Eq.(\ref{eq:motion}).

  Before leaving this point, we would like to elaborate more on the symmetrization introduced in the above set of equations. Note that there is no claim that this is the standard way to project the Hermitian components of a a differential vector operator as in Eqs.(\ref{P_a Hermitianb}) and (\ref{P_a Hermitianc}), nor to construct the full operator from these compinents as in Eq.(\ref{symmetric vector p_a}). What the symmetrization says is that  $\vec{p}_a^{\,H}$; the Hermitian momentum vector component  along $\vec{u}^{a}$ is given by Eq.(\ref{symmetric vector p_a}), and $p_a^{H}$; the component of momentum along the same direction is given by  
Eq.(\ref{P_a Hermitiana}), with the latter, when found according to Eqs.(\ref{P_a Hermitianb}) and (\ref{P_a Hermitianc}),leading to the Hermitian $\vec{p}_a^{\,H}$ and to the correct full momentum operator, Eq.(\ref{symmetric gradient}). This can be taken as the rationale for symmetrizing in the momenta $p_a^{H}$. One can actually provide one more rationale for defining the components of $\vec{p}_a^{\,H}$ as in Eqs.(\ref{P_a Hermitian}): One can say that the vector operator $\vec{p}_a^{\,H}$ is defined through its action on an arbitrary quantum states $\Psi$ , and so we find its components by finding the components of $\vec{p}_a^{\,H}\Psi$, which upon using the form given in Eq.(\ref{vector p hermitian}) reads:
\begin{equation}\label{p on psi}
\begin{aligned}
\vec{p}_b^{\,H}\Psi=&-i \hbar(\vec{u}^b \partial_b+\frac{1}{2 \sqrt{g}} \partial_b(\sqrt{g} \vec{u}^b))\Psi\\
=&-i\hbar(\vec{u}^b (\partial_b\Psi)+\frac{1}{2\sqrt{g}}\left(\partial_{b} \sqrt{g}\right) \vec{u}^{b}\Psi+(\partial_{b} \vec{u}^{b})\Psi)
\end{aligned}
\end{equation}
 Now, with all the derivatives "stuck" to $\Psi$, it does not make difference whether one contracts with $\vec{u}_a$ from right or left!. The result is :
 \begin{equation}\label{}
p_a^{H}\Psi=\vec{u}_a\cdot\vec{p}_b^{\,H}\Psi=\vec{p}_b^{\,H}\Psi\cdot\vec{u}_a= -i \hbar\left(\partial_a+\frac{\Gamma_{a k}^{k}}{2} \right)\Psi
 \end{equation}
 which is just what the expression we get  by symmetrization as in Eqs.(\ref{P_a Hermitian}).
 
 The canonical commutation relations (CCR's) are satisfied for the definition, Eq. ( \ref{P_a Hermitian}), of $p_{a}^H$. Indeed:

\begin{equation}\label{}
\begin{aligned}
& {\left[q^{a}, p_{b}^H\right]=i \hbar \delta^a_b} \\
& {\left[p_{a}^H, p_b^H\right]=0 .}
\end{aligned}
\end{equation}
The last relation above might seem not obvious at first, with $p^H_a=-i \hbar\left(\partial_a+\frac{1}{2} \Gamma_{k a}^{k}\right)$, as the commutator will involve the derivatives of $\Gamma_{k a}^{k}$. Recalling  $\frac{1}{2} \Gamma_{k a}^{k}=\frac{1}{2 \sqrt{g}} \partial_a(\sqrt{g})=\frac{1}{2} \partial_a\left(\ln \sqrt{g}\right)$ the result follows. While the above CCR's are just the same as the Cartesian ones, we can readily compute two commutators that are entirely different from their corresponding (vanishing)  Cartesian ones; the commutator of the canonical momenta with the full momentum  operator, and with the vector components of momentum operator, respectively,  do not vanish:
\begin{equation}\label{non-vanishing CR1}
 [p_a^H,\vec{p}\,]=\hbar^2\vec{u}^{\,j}(\Gamma_{aj}^i\partial_i+\frac{1}{2}\partial_j\Gamma_{ak}^k)
\end{equation}
\begin{equation}\label{non-vanishing CR2}
 [p_a^H,\vec{p}^{\,H}_b]=\frac{i\hbar}{2}\Gamma_{ak}^b (\vec{u}^{\,k} p^H_b+ p^H_b \vec{u}^{\,k}) 
\end{equation}
These results, mean that, unlike in Cartesian coordinates, in GCC's it might not, generally, be possible to find mutual  eigenstates of $\vec{p},\vec{p}^{\,H}_a $, and $p_a^H$.  It is worth mentioning here that the work \cite{Leaf FP} attempts to find eigenstates of $p_a^H$ that are not eigenstates of $\vec{p}$, the interested reader might refer to the article. The above non-commutativity among the operators can also be understood from the perspective of translational invariance and the fact that, when expressed in GCC's in a Euclidean space, the full momentum operator is indeed the generator of space translation, while the individual components are, generally, not \cite{shikakhwa new}.

%-----------------------------------------------------------------

\section{Discussion and conclusions}
In conclusion, we have developed a simple, natural and unified approach to construct the Hermitian vector and canonical components of the momentum operator in GCC's. We have noted that upon identifying the momentum as (mass times) the velocity operator, then with the latter found by applying the Heisenberg equation to the position vector, the resulting expression returns ( $- i\hbar$ times) the gradient plus a zero-values term, which when distributed among the components of the gradient renders each of them equal to the Hermitian  vector component, and when symmetrized in the base vectors, its components are the canonical momentum operators. Thus, we have an algorithm for constructing this operator in any GCC's by first constructing the vector components as per Eq.(\ref{polar vector p's lumped}) (Eq.(\ref{vector p hermitian})), then extracting the canonical components either by symmetrizing as per Eq.(\ref{symmetric polar})  or by projecting these out as per Eq.(\ref{projected canonical polar p's}) (Eq.(\ref{P_a Hermitian})). \\

%A word of warning is due here: These components that we call Hermitian are not necessarily physical or measurable quantities. As we noted in the text, our definition of Hermicity does not guarantee the more general self-adjointness that needs to be met for measurability. To check this, one has to consider each specific coordinate system separately. Discussing this point further is beyond the scope of this work. The interested reader can consult \cite{Paz,Kato}, for example.
 
\section*{Acknowledgments}
The author thanks the colleagues at the Department of Physics of METU; Prof. Dr. S.Turgut and Prof. Dr.T.Alievfor very helpful and stimulating discussions.  He also thanks the department for giving him the opportunity to teach the elective course that triggered the initiation of this work.
%\section*{Author Declaration}
The author has no conflicts of interest to declare.
\section*{Appendix}
\subsection{A.1 Derivation of Eq.(\ref{mv polar})}
\begin{align}\nonumber
\frac{m}{i\hbar}\,[\vec r,H]
&= \frac{i\hbar}{2}
\left[
r\er,
\partial_r^2 + \frac{1}{r}\partial_r + \frac{1}{r^2}\partial_\phi^2
\right]
\end{align}

\noindent
We decompose this commutator into three parts:
\begin{equation}\nonumber
A+B+C ,
\end{equation}
with

\begin{align}\nonumber
A
&= \frac{i\hbar}{2}\,[r\er,\partial_r^2]
= \frac{i\hbar}{2}\,\er
\left(
[r,\partial_r]\partial_r + \partial_r[r,\partial_r]
\right)
\nonumber\\
&= -\,i\hbar\,\er\,\partial_r ,\nonumber
\end{align}

\begin{align}\nonumber
B
&= \frac{i\hbar}{2r}\,[r\er,\partial_r]
= -\,\frac{i\hbar}{2r}\,\er ,\nonumber
\end{align}

\begin{align}\nonumber
C
&= \frac{i\hbar}{2r^2}\,r\,[\er,\partial_\phi^2]
= \frac{i\hbar}{2r}
\left(
[\er,\partial_\phi]\partial_\phi
+ \partial_\phi[\er,\partial_\phi]
\right)
\nonumber\\
&= -\,\frac{i\hbar}{r}\,\ephi\,\partial_\phi
+ \frac{i\hbar}{2r}\,\er .\nonumber
\end{align}

\noindent
Adding \(A\), \(B\), and \(C\), expressing the unit vectors in terms of the contravariant base vectors as in Eq.(\ref{cv polar base vectors}), and re-arranging terms , we obtain
\begin{align}\nonumber
\frac{m}{i\hbar}\,[\vec r,H]
&= -\,i\hbar
\left[
\vec{u}^r\,\partial_r
+ \frac{1}{r}\,\vec{u}^\phi\,\partial_\phi
+ \frac{1}{2r}
\left(
\partial_r(r\vec{u}^r)
+ \partial_\phi (r\vec{u}^\phi)
\right)
\right]
\end{align}
Recalling that $\sqrt{g}=r$, Eq.(\ref{mv polar}) follows.

\subsection*{A.2 Derivation of Eq.(\ref{appendix result}); $p_a^{H}=\frac{1}{2}\left(\vec{u}_a \cdot \vec{p}^{\,H}_b+\vec{p}^{\,H}_b\cdot \vec{u}_a\right)$:}
It is easiest to start with the form of $\vec{p}^{\,H}_b$ given in the first line of Eq.(\ref{symmetric vector p_a}).Using the identities of Christoffel symbols given in Eq.(\ref{CS identities})  we then have:
\begin{equation*}
\begin{aligned}
\frac{1}{2}\left(\vec{u}_a \cdot \vec{p}^{\,H}_b+\vec{p}^{\,H}_b\cdot \vec{u}_a\right)=&\frac{-i\hbar}{2}(\vec{u}_a\cdot(\vec{u}^b\partial_b+\frac{1}{2}\Gamma_{bk}^k\vec{u}^b+
\frac{1}{2}(\partial_b\vec{u}^b))\\
+&\frac{-i\hbar}{2}((\vec{u}^b\partial_b+\frac{1}{2}\Gamma_{bk}^k\vec{u}^b+
\frac{1}{2}(\partial_b\vec{u}^b))\cdot\vec{u}_a)
\end{aligned}
\end{equation*}
Expressing the $(\partial_b\vec{u}^b)$ in terms of the Christoffel symbols as per Eq.(\ref{CS identities}),  contracting with $\vec{u}_a$'s and noting that $\vec{u}^b\cdot\vec{u}_a=\delta_a^b$...etc.  we get:
\begin{equation*}
\begin{aligned}
\frac{1}{2}\left(\vec{u}_a \cdot \vec{p}^{\,H}_b+\vec{p}^{\,H}_b\cdot \vec{u}_a\right)=&\frac{-i\hbar}{2}(2\partial_a+\Gamma_{ak}^k)\\
+&\frac{-i\hbar}{2}(-\frac{1}{2}\Gamma_{bk}^b\vec{u}_a\cdot\vec{u}^k+\Gamma_{ba}^k\vec{u}^b\cdot\vec{u}_k
-\frac{1}{2}\Gamma_{bk}^b\vec{u}^k\cdot\vec{u}_a)
\end{aligned}
\end{equation*}
 Upon contracting the base vectors, the terms in the second line above cancel out leaving :
 
 \begin{equation*}
  \frac{1}{2}\left(\vec{u}_a \cdot \vec{p}^{\,H}_b+\vec{p}^{\,H}_b\cdot \vec{u}_a\right)=-i\hbar(\partial_a+\frac{1}{2}\Gamma_{ak}^k)=p_a^{H}
 \end{equation*}
 
 \subsection*{A.3 Results for the Components of Momentum in Spherical Coordinates}
% =========================
% Appendix: Spherical polar coordinates (answers)
% =========================

We use spherical polar coordinates $(q^1,q^2,q^3)=(r,\theta,\phi)$ with
\[
x=r\sin\theta\cos\phi,\qquad
y=r\sin\theta\sin\phi,\qquad
z=r\cos\theta,
\]
where $r\in[0,\infty)$, $\theta\in[0,\pi]$ and $\phi\in[0,2\pi)$.

The spherical unit vectors are denoted by $\er$, $\etheta$ and $\ephi$.
The covariant base vectors $\vec u_i=\partial_i\vec r$ are
\[
\vec u_r=\er,\qquad
\vec u_\theta=r\,\etheta,\qquad
\vec u_\phi=r\sin\theta\,\ephi.
\]

The metric and its determinant are
\[
(g_{ij})=\mathrm{diag}(1,r^2,r^2\sin^2\theta),
\qquad
(g^{ij})=\mathrm{diag}\!\left(1,\frac{1}{r^2},\frac{1}{r^2\sin^2\theta}\right),
\qquad
\sqrt{g}=r^2\sin\theta.
\]

The contravariant base vectors $\vec u^{\,i}=g^{ij}\vec u_j$ are
\[
\vec u^{\,r}=\er,\qquad
\vec u^{\,\theta}=\frac{1}{r}\,\etheta,\qquad
\vec u^{\,\phi}=\frac{1}{r\sin\theta}\,\ephi.
\]

The Laplacian in spherical coordinates reads
\[
\nabla^2
=\frac{1}{r^2}\partial_r\!\left(r^2\partial_r\right)
+\frac{1}{r^2\sin\theta}\partial_\theta\!\left(\sin\theta\,\partial_\theta\right)
+\frac{1}{r^2\sin^2\theta}\partial_\phi^2 .
\]

Using the expressions above, one verifies the identity
\[
\partial_r\!\left(\sqrt{g}\,\vec u^{\,r}\right)
+\partial_\theta\!\left(\sqrt{g}\,\vec u^{\,\theta}\right)
+\partial_\phi\!\left(\sqrt{g}\,\vec u^{\,\phi}\right)
=\vec 0,
\]
or equivalently,
\[
\frac{1}{\sqrt{g}}\,\partial_i\!\left(\sqrt{g}\,\vec u^{\,i}\right)=\vec 0 .
\]

For the Hamiltonian $H=-\frac{\hbar^2}{2m}\nabla^2+V(\vec r)$, the Heisenberg equation yields
\[
m\frac{d\vec r}{dt}\,
=-i\hbar\left[
\vec u^{\,r}\partial_r
+\vec u^{\,\theta}\partial_\theta
+\vec u^{\,\phi}\partial_\phi
+\frac{1}{2\sqrt{g}}\Big(
\partial_r(\sqrt{g}\,\vec u^{\,r})
+\partial_\theta(\sqrt{g}\,\vec u^{\,\theta})
+\partial_\phi(\sqrt{g}\,\vec u^{\,\phi})
\Big)
\right] .
\]
Since the last term vanishes identically, one recovers $m\,\dot{\vec r}=\vec p$ with
$\vec p=-i\hbar\,\vec u^{\,i}\partial_i$.

The Hermitian vector momentum components are therefore
\[
\vec p^{\,H}_i
=-i\hbar\left(
\vec u^{\,i}\partial_i
+\frac{1}{2\sqrt{g}}\partial_i(\sqrt{g}\,\vec u^{\,i})
\right),
\qquad i=r,\theta,\phi .
\]

In terms of spherical unit vectors, these read explicitly
\[
\vec p^{\,H}_r
=-i\hbar\left(\er\,\partial_r+\frac{1}{r}\er\right),
\]
\[
\vec p^{\,H}_\theta
=-i\hbar\left(
\frac{1}{r}\etheta\,\partial_\theta
+\frac{1}{2r}\cot\theta\,\etheta
-\frac{1}{2r}\er
\right),
\]
\[
\vec p^{\,H}_\phi
=-i\hbar\left(
\frac{1}{r\sin\theta}\ephi\,\partial_\phi
-\frac{1}{2r}\er
-\frac{1}{2r}\cot\theta\,\etheta
\right).
\]

The sum $\vec p^{\,H}_r+\vec p^{\,H}_\theta+\vec p^{\,H}_\phi$ reproduces the full
momentum operator $\vec p=-i\hbar\,\vec u^{\,i}\partial_i$. To find the canonical components, we contract symmetrically with the respective base vectors

\[
p_r^{\,H}=\frac12\left(\vec u_r\cdot\vec p_r^{\,H}+\vec p_r^{\,H}\cdot\vec u_r\right),\quad
p_\theta^{\,H}=\frac12\left(\vec u_\theta\cdot\vec p_\theta^{\,H}+\vec p_\theta^{\,H}\cdot\vec u_\theta\right),\quad
p_\phi^{\,H}=\frac12\left(\vec u_\phi\cdot\vec p_\phi^{\,H}+\vec p_\phi^{\,H}\cdot\vec u_\phi\right).
\]
giving,
\[
p_r^{\,H}=-i\hbar\left(\partial_r+\frac{1}{r}\right),\qquad
p_\theta^{\,H}=-i\hbar\left(\partial_\theta+\frac12\cot\theta\right),\qquad
p_\phi^{\,H}=-i\hbar\,\partial_\phi.
\]


\begin{thebibliography}{99}

\bibitem{Griffiths}
D.~J.~Griffiths and D.~F.~Schroeter,
{\it Introduction to Quantum Mechanics}, 3rd ed. (Cambridge University Press, Cambridge, 2018).

\bibitem{Gaziorowicz}
S.~Gasiorowicz,
{\it Quantum Physics}, 3rd ed. (Wiley, Hoboken, NJ, 2003).

\bibitem{sakurai}
J.~J.~Sakurai and J.~Napolitano,
{\it Modern Quantum Mechanics}, 2nd ed. (Addison–Wesley, San Francisco, 2010).

\bibitem{shikakhwa_2019physicaE}
M.~S.~Shikakhwa,
``The quantum centripetal force on a free particle confined to the surface of a sphere and a cylinder,''
Physica E {\bf 108}, 249 (2019).

\bibitem{shikakhwa2018_commTP}
M.~S.~Shikakhwa,
``Symmetric surface momentum and centripetal force for a particle on a curved surface,''
Commun. Theor. Phys. {\bf 70}, 263 (2018).

\bibitem{shikakhwa2022_EPJPlus}
M.~S.~Shikakhwa and N.~Chair,
``Constructing Hermitian Hamiltonians for spin‑zero neutral and charged particles on a curved surface: physical approach,''
Eur. Phys. J. Plus {\bf 137}, 560 (2022).

\bibitem{shikakhwa2024_curve}
M.~S.~Shikakhwa and N.~Chair,
``Hamiltonian, geometric momentum and force operators for a spin‑zero particle on a curve: physical approach,''
Eur. Phys. J. Plus {\bf 139}, 559 (2024).

\bibitem{Liu} Q. H. Liu, L. H. Tang, and D. M. Xun, ``Geometric momentum: The proper momentum for a free particle on a two-dimensional sphere,'' Physical Review A {\bf 84},  042101 (2011).
    
\bibitem{quesne2004}
C.~Quesne and V.~M.~Tkachuk, 
``Deformed algebras, position‑dependent effective masses and curved spaces: An exactly solvable Coulomb problem,''
J. Phys. A: Math. Gen. {\bf 37}, 4267 (2004); arXiv:math‑ph/0403047.

\bibitem{merzbacher}
E.~Merzbacher,
{\it Quantum Mechanics}, 3rd ed. (Wiley, New York, 1998).

\bibitem{park}
D.~Park,
{\it Introduction to the Quantum Theory}, 3rd ed. (McGraw–Hill, New York, 1992).

\bibitem{DeWitt1}
B.~S.~DeWitt, 
``Point transformations in quantum mechanics,''
Phys. Rev. {\bf 85}, 653 (1952).

\bibitem{DeWitt2}
B.~S.~DeWitt, 
``Dynamical theory in curved spaces. I. A review of the classical and quantum action principles,''
Rev. Mod. Phys. {\bf 29}, 377 (1957).

\bibitem{Cade}
Cade R. Curvilinear momenta in quantum mechanics. Mathematical Proceedings of the Cambridge Philosophical Society. 1951;47(2):451-453. doi:10.1017/S0305004100026803

\bibitem{gruber1971}
G.~R.~Gruber, 
``Quantization in generalized coordinates,''
Found. Phys. {\bf 1}, 227 (1971).

\bibitem{Leaf}
B.~Leaf, 
``Momentum Operators in Curvilinear Coordinates,''
Am. J. Phys. {\bf 39}, 1199 (1971).

\bibitem{Leaf FP}
B.~Leaf, 
``Curvilinear coordinate and momentum operators in configuration representation,''
Found. Phys. {\bf 9}, 575 (1979).

%\bibitem{gruber1972a}
%G.~R.~Gruber, 
%``On quantum mechanical operators in generalized coordinates,''
%Am. J. Phys. {\bf 40}, 1537 (1972).

%\bibitem{gruber1972b}
%G.~R.~Gruber, 
%``Quantization in generalized coordinates—II,''
%Int. J. Theor. Phys. {\bf 6}, 31 (1972).


\bibitem{auckland}
Piaras Kelly, 
\emph{Mechanics Lecture Notes Part III: Foundations of Continuum Mechanics}, 
Department of Engineering Science, University of Auckland, New Zealand, 2015.
Available online: http://homepages.engineering.auckland.ac.nz/~pkel015/SolidMechanicsBooks/Part III/ (accessed July 2025).

\bibitem{Boas}
M. L. Boas, Mathematical Methods in the Physical Sciences, 3rd ed. (Wiley, Hoboken, NJ, 2006).

\bibitem{Arfken}
G. B. Arfken, H. J. Weber, and F. E. Harris, Mathematical Methods for Physicists, 7th ed. (Academic Press, Waltham, MA, 2013).

\bibitem{Dennery}
P.Dennery and a.Krzywicki,
{\it Mathematics for Physicists},  (Dover publication Company, New York, 1995)

%\bibitem{Gruber73}
%G.~R.~Gruber, 
%``Quantization in generalized coordinates—III-lagrangian Formulation,''
%Int. J. Theor. Phys. {\bf 7}, 253 (1973).
\bibitem {shikakhwa new} M.Shikakhwa (2025) to be published.


\end{thebibliography}
\end{document}